\documentclass{amsart}
\usepackage{mathrsfs,amssymb,amsmath,amsfonts}
\usepackage{verbatim}
\usepackage{graphicx,color}
\usepackage{hyperref}

\begin{document}
\title[The background as a quantum observable]{The background as a quantum observable: Einstein's hole argument in a quasiclassical context}
\author{I. Schmelzer}
\email{ilja.schmelzer@gmail.com}
\urladdr{ilja-schmelzer.de}
\thanks{Berlin, Germany}
\sloppypar

\begin{abstract}
I consider a thought experiment measuring the decoherence for quasiclassical superpositions of gravitational field. A version of the hole argument allows to prove that a covariant (background-free) theory is completely unable to define the outcome of this experiment and is therefore not viable. Instead, the results of experiments of this type allow to reconstruct a common background shared by all superposed gravitational fields.
\end{abstract}

\maketitle

\theoremstyle{plain}
\newtheorem{theorem}{Theorem}
\newtheorem{thesis}{Thesis}
\newtheorem{principle}{Principle}
\newtheorem{definition}{Definition}
\newtheorem{postulate}{Postulate}

\newcommand{\pd}{\partial}

\newcommand{\R}{\mbox{$\mathbb{R}$}}
\renewcommand{\H}{\mbox{$\mathcal{H}$}}
\newcommand{\x}{\mathbf{x}} 
\newcommand{\xt}{\mbox{$(\x,t)$}}
\newcommand{\qk}[1]{{\mbox{$|#1\rangle$}}}		
\newcommand{\lk}[1]{\qk{\phi_{#1}}}			
\newcommand{\hk}[1]{\qk{\psi_{#1}}}			
\newcommand{\gk}[1]{\qk{\gc{#1}}}			
\newcommand{\qb}[1]{{\mbox{$\langle{#1}|$}}}		
\newcommand{\lb}[1]{\qb{\phi_{#1}}}			
\newcommand{\hb}[1]{\qb{\psi_{#1}}}			
\newcommand{\gb}[1]{\qb{\gc{#1}}}			
\newcommand{\gc}[1]{\mbox{$g^{#1}_{\mu\nu}$}}		
\newcommand{\gx}[1]{\gc{#1}\xt}				
\newcommand{\lw}[1]{\mbox{$\phi_{#1}$}}			
\newcommand{\hw}[1]{\mbox{$\psi_{#1}$}}			
\newcommand{\lx}[1]{\lw{#1}\xt}			
\newcommand{\hx}[1]{\hw{#1}\xt}			
\newcommand{\ps}[1]{\qk{\phi_{#1}}}		
\newcommand{\psn}[1]{\lk{_{#1}}}		
\newcommand{\pro}{\mbox{$\langle \phi_1|\phi_2\rangle$}}
\newcommand{\abs}[1]{\mbox{$|#1|$}}
\newcommand{\iabs}[1]{\left|#1\right|}

\newcommand{\unit}{\mbox{$\hat{1}$}}
\renewcommand{\a}{\alpha}		
\renewcommand{\o}{\theta}		
\newcommand{\ao}{\abs{\o}}		
\newcommand{\po}{e^{i\a}}		
\newcommand{\so}{\mbox{$\hat{\rho}_\o$}}	
\newcommand{\oo}{\mbox{$\o$}}		
\newcommand{\Sch}{Schr\"{o}dinger\/ }

\section{Introduction}

The quantization of general relativity is known to be extremely difficult: More than eighty years after the creation of the two theories themself the problem is yet not solved in a ``satisfactory'' way. Here, with ``satisfactory'' in scare quotes we mean a quite popular requirement -- that the theory should be background-independent. This property is considered by many physicists (for example Smolin \cite{Smolin} or Thiemann \cite{Thiemann}) as an extremely important property of classical general relativity which should be preserved in its quantum version. 

The aim of this paper is to show that the problem is not only difficult -- a background-free theory of quantum gravity is impossible. We propose a thought experiment such that its result cannot be predicted by a background-free theory. This sounds negative, but the message is a positive one. It follows that the quantum interference effects we consider give us new, non-classical observational possibilities. The results of these experiments not only depend on the background, but contain at least in principle sufficient information to recover the background. Roughly speaking, the background becomes observable in quantum gravity.

The experiment is a partial position measurement for a sufficiently heavy particle (which is in a superpositional state) by gravitational interaction with a lightweight particle. The question is if the gravitational interaction leads to decoherence or not, and the degree of decoherence. For this purpose, we can ignore the state of the test particle, and it is sufficient to observe the resulting state of the heavy particle.

With ``electromagnetic'' instead of ``gravitational'', this experiment can be easily realized. Unfortunately the gravitational field is simply too weak to give nontrivial observable results. But it is straightforward to compute the result in Newtonian quantum gravity (another word for \Sch theory with Newtonian interaction potential). We start with the two-particle state
\begin{equation}\label{e:initial}
 \Psi_{in} = \frac{1}{\sqrt{2}}(\hk{1}+\hk{2})\lk{0},
\end{equation}
where the states $\hk{i}$ of the heavy particle are localized near different points $x_i$ so that $\hx{i} \approx \delta(\x-\x_i)$, and \qk{\phi_0} denoted the initial state of the lightweight test particle. Because of the mass difference, the Newtonian gravitational interaction will leave the basic states of the heavy particle \hk{i}\/ approximately unchanged, while the resulting state of the lightweight particle changes. Nonetheless, if the heavy particle is in one of the localized basic states \hk{i}, the resulting state will remain approximately in a product state, with the lightweight particle in the resulting state \lk{i}\/ which (if the difference in the gravitational forces is large enough) depends on the position of the heavy particle $x_i$. For the superpositional initial state \eqref{e:initial} this leads to a resulting two-particle state
\begin{equation}
 \Psi_{out} = \frac{1}{\sqrt{2}}(\hk{1}\lk{1}+\hk{2}\lk{2})
\end{equation}
The ignorance of the state of the lightweight test particle leads to a density matrix state \so\/ for the heavy particle defined by
\begin{equation}\label{e:rho}
\so = \frac12\Bigl((\sum\limits_i\hk{i}\hb{i}) + \o\hk{2}\hb{1} + \overline{\o}\hk{1}\hb{2}\Bigr)
\end{equation}
where the complex number $\o$ is defined by the scalar product
\begin{equation}\label{e:product}
\o = \pro = \int \overline{\phi}_{1}(\x)\phi_{2}(\x) d^3\x.
\end{equation}
Observing the remaining interference pattern defined by this state of the heavy particle, we can measure \oo. The remaining degree of interference is defined by $\ao$. For $\ao=0$ no interference pattern is observable, the interference has been completely lost because of decoherence. Else, one can also observe a possible phase shift of the interference pattern caused by the gravitational interaction, which defines the phase of \oo.

One may want to compute relativistic corrections. For the heavy particle localized exactly in $x_i$ semiclassical gravity allows to compute such corrections. But superpositions of states described by different gravitational fields are already outside the domain of applicability of the semiclassical approximation.

Nonetheless, one can make a naive attempt. One can use semiclassical gravity to compute general-relativistic corrections for the wave function \lx{i}, and then simply apply \eqref{e:product}. That means, instead of computing the \lk{i}\/ considering their Newtonian interactions with the \hk{i}\/ localized in $x_i$, we compute the \lk{i}\/ considering the movement of a single particle (ignoring as a first approximation particle creation and destruction and other semiclassical subtleties) on a curved background \gx{i}, which is defined as the GR solution with the heavy (classical) particle localized in $x_i$. Then we simply put the resulting wave functions \lw{i}\/ into formula \eqref{e:product} and have obtained a prediction for \oo.

Unfortunately this naive computation does not work. The problem is the choice of the coordinates. What is the system of coordinates $\x$ to be used in \eqref{e:product}?

To see the problem let's remember that GR defines the solutions \gx{}\/ only modulo arbitrary changes of coordinates. This remains true also in semiclassical GR. In this case, the change of coordinates leads to a nontrivial transformation of the fields moving on the background too. If we restrict ourself for simplicity to changes of coordinates which leave time $t$ unchanged, wave functions will transform like square roots of densities. Thus, \lw{}\/ would have to be replaced by another wave function $\phi'$. As long as we consider only one background, this does not cause problems. But in \eqref{e:product} two different background fields \gx{i}\/ are involved, and there is a freedom of choice of coordinates on each of them. Thus, nothing prevents us from changing the system of coordinates for \lx{2}, \gx{2}, leaving the coordinates on \lx{1}, \gx{1}\/ unchanged. It follows that \lx{2}\/ has to be replaced by a transformed wave function $\phi'_2\xt$, but \lx{1}\/ remains fixed. But the scalar product \eqref{e:product} does not remain invariant:
\begin{equation}
\int \overline{\phi}_{1}(\x)\phi_{2}(\x) d^3\x \neq \int \overline{\phi}_{1}(\x)\phi'_{2}(\x) d^3\x.
\end{equation}

Which value defines the observable \oo\/ in \eqref{e:product}? This is the problem we consider in this paper.

One could think that we have to use in \eqref{e:product} a system of coordinates for \lw{2}\/ which in some sense ``corresponds'' to the system of coordinates used for \lw{1}. Unfortunately, GR does not define such ``corresponding'' systems of coordinates for different solutions. The systems of coordinates of different solutions are in no way connected with each other. In principle, different solutions may live even on topologically different manifolds, so that no continuous correspondence relation between the events of the different solutions exists. But even if the manifolds are, in above cases, diffeomorph to $\R^4$, to define a scalar product we would have to fix some diffeomorphism between them. But GR does not fix any such diffeomorphism between different solutions. The scalar product \eqref{e:product} is simply undefined in GR.

Another natural reaction is to blame quantum strangeness for this problem. We need a full quantum theory of gravity to make a prediction, but we do not have such a theory yet. But do we really need it? Quantum strangeness has not prevented us from computing the prediction in Newtonian quantum gravity. Even more, there is a simple class of classical theories of gravity such that quantum strangeness does not prevent us from computing the prediction using the same formula \eqref{e:product} as in Newtonian quantum gravity, following the naive scheme considered above. All we need to apply \eqref{e:product} is a \emph{background}. Once we have a common background shared by all gravitational fields, the problem disappears. The system of coordinates $\x$, to be used in \eqref{e:product}, is simply the system of coordinates of the background. The background gives exactly what we need -- corresponding systems of coordinates for different solutions.

Thus, there is no reason to blame quantum strangeness for the failure of GR. The competing theories of gravity with background, faced with the same quantum strangeness, have no problem at all with computing a prediction for our simple thought experiment. The rules of the scientific method for thought experiments seem sufficiently clear about what to do in this case. Popper's criterion of empirical content may be applied also to predictions about thought experiments -- the theory which makes more predictions has to be preferred. Theories with background make predictions, GR not, thus, theories with background have to be preferred.

But this is hardly a sufficient argument for a community which thinks that ``background independence is \ldots \textbf{The Key Feature} of quantum gravity'' \cite{Thiemann} (original emphasis). If combined with the idea that ``[t]he whole formalism of ordinary QFT heavily relies on this background structure and collapses to nothing when it is missing''  \cite{Thiemann} and  that ``[n]o matter how one deals with this issue \ldots one has to invent something drastically new in order to quantize the gravitational field''  \cite{Thiemann}, no particular mathematical problem of background-independent theories seems sufficient.

But is this argument strong enough to destroy the strong support for GR and the philosophical idea of background freedom? The aim of this paper is to consider this question in more detail. We find that the problem is not only a GR problem, but one of the concept of background freedom. We prove an impossibility theorem for a whole class of theories with background freedom which is more general than GR. It contains, in particular, also theories which solve the problem of time in a trivial way, by the introduction of a preferred foliation, but are spatially background-free, that means, have active diffeomorphism invariance for diffeomorphisms which preserve the preferred foliation.

We also look at the situation from the other side. Given a theory with background, we can compute the \oo. But can we, given the \oo\/ for all imaginable experiments of this type, recover the background? We find a positive answer. In other words, with the thought experiment presented here, the background becomes in principle observable.

Given that the average reader may not know them, we give a very short introduction into a few theories of gravity with background we consider most promising. And we discuss, in a lot of (hopefully not boring) detail, lots of possible objections.

\section{Definitions}

The aim of the paper is to prove an impossibility result for a certain, quite general class of quantum theories of gravity. But we do not fix properties of the theory itself. Instead, we fix only properties of certain approximations. We consider different approximations, and sometimes have to switch between these different approximations. Moreover, we want to cover different theories of gravity, which differ already in their classical approximations. Because this collection of different approximations of different theories may become confusing, it seems useful to give here an overview and a description of the precise meaning of the terms we use to describe these theories and approximations. By the way, we introduce some denotations of the objects of these theories and approximations.

\begin{itemize}
\item First, there is \emph{classical gravity.} This term means non-quantum, not non-relativistic. We consider here only metric theories of gravity, thus, the gravitational field is assumed to be described by a spacetime metric \gx{}. There are three interesting subclasses of these theories: First, \emph{completely background-free theories}. This is, essentially, general relativity. Then, \emph{theories with background}. Here, the important examples are bimetric theories with a Minkowski background and theories which assume a classical Newtonian spacetime $\R^3\times\R$ as a background. The key property of these theories is that for any event $\xt$ of one solution there exists a corresponding event (denoted by the same $\xt$) on each other solution, provided that such an identification is given for the initial values $t=t_0$ and boundary conditions. The third class are \emph{spatially background-free theories}. Here, the property of background freedom is restricted to diffeomorphisms which preserve a given time coordinate $t$: The theory should be covariant for arbitrary coordinate transformations of the form $t'=t$, $\x'=\x'(\x,t)$. A generally covariant theory is, of course, covariant for this restricted class of diffeomorphisms too, whatever the choice of the time coordinate. Thus, completely background-free theories are also spatially background-free.

To define a solution of classical gravity we use the denotation \gx{}. But let's note here that the restriction to metric theories of gravity is not essential, the argumentation would not change much if gravity would be described by other fields -- only the denotations would become more uncommon. The denotation \gx{i} is used for a particular solution of classical gravity which contains a heavy point particle at the position $x_i$.

\item Next, there is the \emph{semiclassical approximation}. The gravitational field in the semiclassical approximation is described by a solution of the corresponding classical theory of gravity \gx{}. On this background, quantum theory is developed. So, the gravitational field influences the quantum fields, but the state of the quantum fields does not influence the gravitational field. We have localized one particle near some points $x_i$ and described it as ``heavy'', but the other one as ``lightweight'' to make this approximation applicable: The heavy particle in a classical position $x_i$ defines a classical gravitational field \gx{i}, which influences the quantum lightweight particle, but the influence of the lightweight particle on the heavy particle in $x_i$ and on the gravitational field \gx{i}\/ created by the heavy particle can be ignored.

Semiclassical gravity is an established theory, usually applied in combination with GR as the corresponding classical theory. But the techniques do not depend on this and may be applied as well if a different classical metric theory of gravity is used. To describe the state of the quantum system on the background  \gx{i}\/ we use the denotation \lk{i}.

\item Then there is the \emph{one-particle approximation of the semiclassical approximation.} In full semiclassical approximation, there are not only effects of particle creation and destruction by the gravitational field, but even the definition of the notion of particles and the vacuum state is problematic. Therefore one usually considers field theories as the quantum theories on curved backgrounds. Nonetheless, there will be some domain (say, sufficiently weak gravitational fields combined with a sufficiently nice time coordinate $t$) where a one-particle approximation is appropriate.

In these cases, the quantum state \lw{}\/ can be described by a single particle wave function denoted by \lx{}. In particular, the wave function of the test particle on the background \gx{i} the wave function will be denoted by \lx{i}.

\item Then there is the approximation we have named here \emph{quasiclassical approximation}. The notion ``quasiclassical'' is widely used in different meanings, so this is a sort of non-standard use, but we have not found a better notion to distinguish this class of approximations.

This class contains all the states described in a semiclassical approximation but also finite superpositions of semiclassical states which correspond to different classical metrics. For a state connected with a classical gravitational field \gx{i}\/ we use the denotation \gk{i}. Together with the quantum state of the lightweight particle \lk{i}\/ on this background this defines the state $\gk{i}\otimes\lk{i}$. Such states will be named ``semiclassical'', and may be superposed with other semiclassical states which may be connected with other classical metrics.

What distinguishs this approximation from full quantum gravity is that we do not care about the quantum uncertainties in the definition of the states \gk{}. In a full quantum theory, operators measuring different components of the metric may not commute, and the quantum state \gk{} may possibly only approximate the classical metric \gx{}\/ at a given moment $t$, similar to coherent states which can only approximate a classical state $\delta(p-p_0)\delta(q-q_0)$. These possible uncertainties are ignored in the quasiclassical approximation. The quasiclassical approximation is conceptually interesting because, on the one hand, the conceptually most important quantum effects -- superpositional states of gravitational fields -- are already present, while, on the other hand, we can yet work with the classical gravitational fields \gx{}. In particular, we can easily define the appropriate notion of covariance or background-freedom in this approximation.

\item Then there is, of course, \emph{Newtonian quantum gravity}. This is simply multi-particle \Sch theory with Newtonian interaction potential, but, despite this simplicity, it is a full (even if only non-relativistic) quantum theory of gravity. In this theory the states of the heavy particle will be denoted by $\hx{i}$ or \hk{i}\/ while states of the lightweight particle will be denoted by  $\lx{i}$ or \lk{i}.

\item Last but not least, there is  \emph{full quantum gravity}. Informally, we use this denotation to describe a quantum theory of gravity which describes also relativistic effects correctly, and assume that it is a quantum theory, not some modification of quantum theory which gives quantum theory only in some limit. But essentially we make only assumptions about properties of various approximations.

\end{itemize}

\section{A thought experiment with a superposition of gravitational fields}\label{s:exp}

To create a superpositional state of the gravitational field is, even if completely unrealistic in practice, conceptually quite easy -- all one has to do is to create a usual superpositional state of some sufficiently heavy source particle. Thus, at least in principle, we do not need more than a usual double slit arrangement, with two slits at two points $x_i$, which creates a superpositional state for a sufficiently heavy source particle, and at least in principle the gravitational field created by this particle is already in a superpositional state.

The resulting superpositional state interacts with its environment, and this interaction leads, if it is sufficiently strong, to decoherence. What we are interested in is the decoherence caused by the gravitational interaction. So we assume that there is some lightweight particle which interacts gravitationally with the superpositional state, and we want to compute the resulting decoherence effect. In practice, this effect is far too small to be important -- the other forces are so much stronger that they cause complete decoherence before the decoherence effect of the gravitational interaction becomes observable. But this is a practical problem, and nothing suggests that the theoretical principles which work for the much stronger electromagnetic interaction become invalid for gravity.

Let's denote the states where the source particle is at a well-defined position $x_i$ with \hk{i}. Then, our initial state is the superposition $\frac{1}{\sqrt{2}}(\hk{1}+\hk{2})$.

\begin{theorem}[the result of partial position measurement]\label{t:measurement}
Assume a pure superpositional state $\frac{1}{\sqrt{2}}(\hk{1}+\hk{2})$ of some system interacts with some part of its environment. After the interaction, the state of the environment is ignored. Assume that the interaction leaves the basic states \hk{i}\/ (approximately) unchanged. Then

\begin{enumerate}
 \item \label{i:density} the resulting state \so\/ of the system has (approximately) the form
\begin{equation}\label{e:rho1}
\so = \frac12\Bigl((\sum\limits_i\hk{i}\hb{i}) + \o\hk{2}\hb{1} + \bar{\o}\hk{1}\hb{2}\Bigr)
\end{equation}
which is uniquely characterized by the complex number $\o$ with \mbox{$\ao\le 1$};

\item \label{i:observable} this number $\o$ is observable.

\item \label{i:pure} If the environment is initially in a pure state \lk{0}, then we have $\o = \pro$, where the \lk{i}\/ are the resulting pure states of the environment if the system is in the state \hk{i}.

\item \label{i:scalar} If, moreover, the environment is defined by a single particle, so that the states \lk{i}\/ are described by single particle wave functions \lx{i}, then
\begin{equation}\label{d:product}
\o = \pro = \int \overline{\lw{1}}\xt \lx{2} d^3x.
\end{equation}
\end{enumerate}
\end{theorem}

\begin{proof}
The interaction is itself is defined by some unitary operator $U$ acting on the tensor product of the Hilbert spaces $\H_S\otimes\H_E$ of the system and the environment. Then, the ignorance of the environment means that all observations are restricted to observations of system observables $O_S$. This is equivalent to a reduction of the complete density matrix on $\H_S\otimes\H_E$ to a density matrix on $\H_S$ alone.

Let's consider at first the case where the environment is initially in a pure state \lk{0}. Then the initial states $\lk{0}\otimes\hk{i}$ remain (because of unitarity of the evolution) to be pure states. But once the states \hk{i}\/ remain finally unchanged, all system observables $O_S$ should give the same result as for the  \hk{i}\/ themself, which happens only if the resulting states are pure product states $\lk{i}\otimes\hk{i}$. Once the evolution is unitary, the initial state $\lk{0}\otimes\frac{1}{\sqrt{2}}(\hk{1}+\hk{2})$ gives
\begin{equation}
\qk{\psi} = \frac{1}{\sqrt{2}}(\lk{1}\otimes\hk{1}+\lk{2}\otimes\hk{2}).
\end{equation}
A measurement of a system observable $O_S$ gives then
\begin{equation}
\qb{\psi} O_S \qk{\psi} = \frac12 \sum\limits_{ij} \lb{i}\unit\lk{j} \hb{i}O_S\hk{j}
\end{equation}
which coinsides with $Tr(\rho O_S)$ for a density matrix of the system of the form
\begin{equation}\label{d:rhoij}
\rho = \sum\limits_{ij} \rho_{ij} \hk{j}\hb{i}; \qquad  \rho_{ij} = \frac12 \lb{i}\unit\lk{j}.
\end{equation}
With this definition of the $\rho_{ij}$, $\rho$ is indeed a density matrix (self-adjoint, trace $1$ and all eigenstates non-negative), with the additional property $\rho_{ii}=\frac12$. This is \eqref{d:product} with $\o = \pro$, so that we have already proven point \ref{i:pure}. Point \ref{i:scalar} is then a trivial consequence.

To prove point \ref{i:density} we have to consider the more general case that the environment is initially already in a mixed state.  This does not give much new, because this can be described by a classical probability combination that the initial state is some \lk{k0}\/ with probability $p_k$. Then the resulting state can also be described as such a combination, and gives a density matrix with the coefficients
\begin{equation}
 \rho_{ij} = \frac12 \sum\limits_k \lb{ki}p_k\lk{kj}.
\end{equation}
This does not change the properties we have mentioned. States connected in this way with a classical solution are named semiclassical states. $\rho$ remains to be a density matrix, and the  additional restriction $\rho_{ii}=\frac12$ remains intact too, so that point \ref{i:density} holds too. 

The coefficients of the density matrix are known to be observable. But to see this in detail, the representation of \so\/ as
\begin{equation}\label{e:rhodecomposition}
\so = (1-\ao)\left(\frac{1}{2}\sum\limits_i\hk{i}\hb{i}\right) + \ao\qk{\psi_\a}\qb{\psi_\a}
\end{equation} 
where
\begin{equation}
\qk{\psi_\a} = \frac{1}{\sqrt{2}}(\hk{1} + e^{i\a}\hk{2}); \qquad \o = \ao e^{i\a}
\end{equation}
may be useful. It shows that with proability $1-\ao$ the heavy particle is in the diagonal state $\frac12\sum\hk{i}\hb{i}$ which does not give any interference, while with probability $\ao$ it is in the pure superpositional state \qk{\psi_\a}, which differs from the initial state only by a (possibly) different phase factor $e^{i\a}$. A standard measurement of an interference pattern allows to measure $\ao$ (the remaining degree of interference) as well as $\po$ if $\ao>0$ (an additional phase shift of the interference pattern in comparison with the initial state).
\end{proof}

Note that the theorem does not refer to any specific properties of the environment and the particular interaction, except that the basic states \hk{i}\/ are left invariant, and this last property has been introduced only for reasons of simplicity. Thus, the formula does not only hold in Newtonian quantum gravity, but in every quantum theory where states of a system and its environment are represented by density operators in a tensor product of Hilbert spaces $\H_S\otimes\H_E$ and evolution is unitary.

Thus, a quite general interaction with some completely unspecified part of the environment, which will be ignored after the interaction, gives an observable effect characterized by some complex number \oo. 

The expression of the scalar product as an integral over $\x$ is, of course, only one possibility. One can measure whatever property of the test particle one likes and, then, ignore the results of this measurement, which gives a representation of the scalar product as an integral over other variables.

\section{The properties of the quasiclassical approximation}

To prove our impossibility theorem we have to make assumptions about the full quantum theory of gravity. The main assumption is that there exists some approximation which we name ``quasiclassical approximation'' which has certain properties.

Now, usually one would prefer not to use approximations -- they are a cause of error and uncertainty. It would be better to use, if possible, the exact theory.

But the logic of impossibility proofs is reverted. To make an impossibility theorem stronger, we have to weaken the assumptions made about the theories. To postulate an exact property of the theory itself is a much stronger restriction than to postulate that the property holds only approximately. The consequence of the latter, weaker assumption in an impossibility theorem is that a much larger class of theories is shown to be invalid. Similarly, impossibility theorems become stronger if they use weaker approximations. Therefore we try to find the weakest approximation of full quantum gravity where the problem appears. We consider this line of argument in more detail later, in the sections \ref{a:whyApproximations}, \ref{a:whyWeakerApproximations}. 

\subsection{The assumption of a special time coordinate}

We assume that, together with the classical solution \gx{}, we have also some special time coordinate $t$.

One may object that this is problematic, because the most interesting case, general relativity, does not define any preferred time coordinate. But it is for this reason that we use the denotation ``special'' instead of ``preferred''. We do not require that the theory \emph{has} a preferred time coordinate. We also do not require that the special time coordinate $t$ is fixed uniquely by the classical theory. We only assume that in the domain of application of the quasiclassical approximation exists some special time coordinate $t$ so that the solution may be described in the form \gx{}, and that this time coordinate has some nice properties specified later: In particular, we want to use the time coordinate to define quantum states \gk{}\/ of the gravitational field which corresponds to some some moment of time $t$ of a classical solution \gx{}. Then, we want to use this time coordinate to define in a one-particle approximation a wave function \lx{}\/ which allows a standard probability interpretation $\rho\xt = \overline{\lw{}}\xt\lw{}\xt$. These are special properties of the time coordinate $t$ which may hold for some time coordinates but not for others. Thus, a time coordinate having these properties is in some sense ``special''.

The time coordinate is, therefore, not a property of the classical theory of gravity, such that the classical theory of gravity has to define, uniquely, the time coordinate $t$.  Its existence is, instead, a particular property of solutions in the domain of applicability of the quasiclassical approximation: Some solutions allow the introduction of a nice time coordinate -- fine, these solutions are inside the domain of applicability of the approximation. Others do not allow such a time coordinate (say, GR solutions with closed causal loops, or extravagant topologies) -- no problem. The approximation is not applicable for this particular solution. But we do not require anywhere that the approximation is applicable to all solutions of the theory. Thus, a theory which allows such a special time coordinate only for some solutions remains to be covered by the impossibility theorem. Another solution allows even several different such nice time coordinates -- no problem too. A particular application of the approximation can use each of them.

\subsection{Theories with preferred time}

But the case of theories which have a preferred time already in the classical theory is, of course, also interesting. To consider such theories is one possible strategy to solve the notorious ``problem of time'' of quantum general relativity. If there exists a fundamental preferred time, the problem of time simply disappears. But what about covariance for purely spatial diffeomorphisms? Even if one gives up to solve the problem of time, and accepts the trivial solution defined by a preferred time, one may like the idea of background freedom so much that one likes to save at least diffeomorphism invariance for purely spatial diffeomorphisms. Or one may consider such a theory as an intermediate step toward a fully background-free theory: First, we find a spatially background-free theory with preferred time, and then we show that all these theories make the same observable predictions, independent of their choice of the preferred time. (This is, essentially, the program of canonical quantum gravity.)

As we will see, these intermediate theories are also covered by our impossibility theorem. The preferred time of such theories has another special property: The theory is covariant for diffeomorphisms which preserve the preferred time, while it is not known if it is covariant for general diffeomorphisms. Again, fully covariant theories like GR are also covered.

But note that in the case of such theories with a ``preferred'' time the ``special'' time coordinate $t$ we have described in the last section is assumed to be identical with the preferred time, even if these notions are conceptually different. The ``preferred'' time is a fundamental property defined already in the classical theory, which exists for all solutions, and is connected with general properties like covariance for diffeomorphisms preserving it. The ``special'' time coordinate is only a particular property of particular solutions which make certain approximations applicable.

This is, again, not problematic. For theories with a preferred time, it means that the domain of applicability of the quasiclassical approximation has a further restriction -- it may be applied only to those solutions where the preferred time may be used also as the special time. For theories without preferred time, this question does not exist. Diffeomorphism invariance is used only for the restricted class of diffeomorphisms preserving special time, but this is sufficient to prove our impossibility theorem. 

\subsection{Semiclassical states}

The first use of the special time coordinate $t$ is the definition of the quantum state \gk{}\/ of the gravitational field given a classical solution \gx{}\/ and a moment of time $t$.

To define such a quantum state, one would assume some connection with the initial value problem of the classical theory of gravity. Thus, one would assume that one has to define at the moment $t$ not only the solution \gx{}\/ itself, but also some number of derivatives in $t$. Moreover, the classical theory can contain some constraints for the initial values which have to be fulfilled. The specification of these initial values and their constraints is something which depends on the classical theory of gravity. Fortunately, we do not have to consider these questions, because the specification by a solution \gx{}\/ together with a moment of time $t$ is sufficiently general, does not depend on the specific constraints of a particular metric theory of gravity.

We also assume that the quantum state \gk{}\/ is, in the approximation, uniquely defined by the classical solution \gx{}\/ and a moment of time $t$.  This will hold only approximately. The initial values will contain field components themself as well as time derivatives of them, and one would not expect that the corresponding operators will commute in the quantum case. If they do not commute, it would be as impossible to represent a metric \gx{}\/ exactly by a quantum state \gk{}\/ as it is impossible to represent the classical state $\delta(q-q_0)\delta(p-p_0)$ by a quantum state which has exactly the position $q_0$ and momentum $p_0$. There are nice replacements, coherent states, which approximate such a state in a sufficiently accurate way, but they are not uniquely defined. One can define coherent states for every choice of accuracies $\Delta p$ and $\Delta q$ which fulfill the uncertainty relations. In a similar way, there may be different states which approximate a given classical metric \gx{}. The semiclassical approximation ignores these subtleties and assumes that there is a unique choice so that every classical metric \gx{}\/ at every moment of time $t$ defines a single quantum state \gk{}\/ in a unique way.

We do not make here any assumptions that different initial values \gx{}\/ have to lead to different states \gk{}. In background-free theories different but equivalent initial values for the classical problem may be described by the same quantum state.

\subsection{One particle approximation}

It is well-known that there is no consistent one-particle theory on a curved background. There are not only effects like particle creation and destruction by the gravitational field, but even the definition of the notion of a particle and the definition of the vacuum state in field theories is problematic on a curved background.

Nonetheless, there is a domain where the one-particle approximation is appropriate -- the domain where one has a preferable choice of the vacuum and particle operators. A standard construction uses a Killing vector field. We have already assumed that we have a special time coordinate and may use this time coordinate for the definition of a preferred notion of a particle.

This is not complicate -- at every point the metric \gx{}, together with the special time coordinate $t$, defines also special vector which describes an observer ``at rest'', whose local time (as defined by local Einstein synchronization) coinsides in linear approximation with the special time $t$. Then, for the definition of the number of particles in some region one can use the response of particle counters which are locally ``at rest''.

But, again, it is not our task to develop some particular approaches to a new theory of gravity, but, instead, to prove that a certain class of theories -- background-free theories -- cannot reach its aims. So we start from the other side. Instead of proposing a particular way to solve this problem, we assume that the theory of gravity has \emph{somehow} solved this problem, without specifying the way. It may be something close to the way suggested in the previous paragraph, or it may be something completely different. The point is that single particle approximations are meaningful in some domain, at least as approximations, thus, a theory which does not allow a single particle approximation seems something completely strange. And, therefore, to assume that there is a domain of quantum gravity where single particle theory is well-defined as an approximation is a completely reasonable and very weak assumption.

Technically, there are two reasons why we prefer to consider the one-particle approximation instead of the (also well understood) semiclassical multi-particle theory:  First, it is more close to Newtonian quantum gravity (multi-particle \Sch theory with Newtonian interaction potential) where particle numbers are conserved, and we can apply the same formula for the scalar product \eqref{e:product} in above theories. But, more important, we follow the principle that the weaker the approximation used, the stronger the impossibility result. So, once the problem exists already in the one-particle approximation, we have to prefer it.

\subsection{The main postulate}

Thus, we assume now that the basic states \hk{1} and \hk{2} are defined by solutions \gx{1}, \gx{2}\/ of some classical metric theory of gravity (not necessarily GR). We restrict ourself to sufficiently weak gravitational fields and a sufficiently plane foliation $t$. In this case, we can ignore effects of particle creation and destruction by the gravitational field and consider a one-particle approximation. The details of the evolution are not relevant -- all we need is that, given the solutions \gx{i}\/ and some initial value for the wave functions \lx{i} at $t_0$ we can compute them for all times $t$. Note that the evolution should not even be unitary, because there will be some probability of particle creation and destruction which will be ignored in our approximation. All we would like to have is summarized in the following postulate:

\begin{postulate}\label{p:main}
The theory contains a subset of sufficiently well-behaved classical solutions  \gx{}\/ named ``weak fields'' such that these weak fields have the following properties: 
\begin{enumerate}
\item \label{i:first} \label{i:time} For each weak field \gx{}\/ exists some corresponding ``special'' time coordinate $t$.
\item \label{i:state} For each weak field \gx{}\/ and each moment of time $t$ the theory defines in an approximately unique way a so-called ``semiclassical state'' \gk{}\/ of the gravitational field.
\item  On the gravitational background defined by semiclassical states \gk{}\/ exists a well-defined single particle approximation so that the state \lk{}\/ of the particle at the moment of time $t$ is defined by a wave function $\lx{}\in\mathbb{C}$. The one-particle wave functions are elements of some Hilbert space $\H_{\gk{}}$ with the scalar product
\begin{equation}\nonumber 
 \pro = \int \overline{\lw{1}}\xt \lw{2}\xt d^3x; \qquad \lw{i}\in H_{\gk{}}.
\end{equation}
\item \label{i:sesqui} For different semiclassical states \gk{1}, \gk{2}\/ exists a non-degenerated sesquilinear form on $\H_{\gk{1}}\times \H_{\gk{2}}$. 

\item \label{i:exp} For superpositional states $\frac{1}{\sqrt{2}}(\gk{1}+\gk{2})$ of semiclassical states, the theory predicts that the outcome $\oo$ of the experiment described in section \ref{s:exp} is defined by this sesquilinear form as
\begin{equation}\nonumber 
\oo = \pro.
\end{equation}
\item \label{i:explicit} The sesquilinear form on $\H_{\gk{1}}\times \H_{\gk{2}}$ is defined by
\begin{equation}\nonumber 
 \pro = \int \overline{\lw{1}}\xt \lx{2} d^3x; \qquad \lw{i}\in H_{\gk{i}}
\end{equation}
\end{enumerate}
\end{postulate}

\section{The classical hole argument}\label{s:hole}

Let's remember Einstein's hole argument \cite{Einstein} and its resolution in classical general relativity. Let $g_{\mu\nu}(x)$ be a solution of the Einstein equations of general relativity (GR). Let's consider some bounded spacetime region $\Sigma$ -- the ``hole''. It is located in the future of the initial time $x^0=t_0$.
\footnote{
In Einstein's version it was located in a region without any material processes. The conclusion was, correspondingly, that the gravitational field cannot be uniquely defined by the distribution of matter. This contradicts Mach's principle, but this is something we have accepted. The gravitational field has its own degrees of freedom, gravitational waves. Thus, this version of the hole argument has only historical interest.
} 
Then we consider a nontrivial smooth diffeomorphism $x' = x'(x)$, which is trivial ($x'(x)=x$) outside $\Sigma$. As a consequence of diffeomorphism invariance, the transformed metric $g'_{\mu\nu}(x')$, after formally replacing $x'$ with $x$, gives a different solution $g'_{\mu\nu}(x)$ of the Einstein equations. It coinsides with $g_{\mu\nu}(x)$ outside $\Sigma$, thus, defines a different solution with the same boundary conditions and the same set of initial data at $x^0=t_0$. Thus, at a first look, it seems that in GR the gravitational field cannot be uniquely defined by whatever set of initial values and boundary conditions, for matter fields as well as the gravitational field.

The GR solution of this problem is that the two solutions $g_{\mu\nu}(x)$ and $g'_{\mu\nu}(x)$, while different as functions of $x$, cannot be distinguished by observation. What naively looks like an observable -- the value $g_{\mu\nu}(x)$ in a point $x$ -- appears to be unobservable: Observables are connected with events, which have to be identified by their relations to other events. For example, the event $x_1$ may be identified by the set of events at $t_0$ which intersect its past light cone. But this same set at $t_0$ defines, on the $g'_{\mu\nu}(x)$, another event $x_1'=x'(x_1)$.

Thus, despite the classical hole argument, in GR all real, physical observables are well-defined by the initial values and boundary conditions. But for the viability of the covariant Einstein equations it is essential that there is no observable $O$ which allows to distinguish the two solutions $g_{\mu\nu}(x)$ and $g'_{\mu\nu}(x)$ connected via the coordinate transformation $x'=x'(x)$ in the hole $\Sigma$.

\section{The impossibility theorem}

In our considerations it is sufficient to consider the hole argument for purely spatial diffeomorphisms $(\x',t') = (\x'(\x,t),t)$, which preserve our fixed foliation of spacetime: Covariance for this subgroup of diffeomorphisms is already sufficient to prove our impossibility theorem. Of course, a fully covariant theory will be covered anyway -- the theory is also covariant for all spatial diffeomorphisms. But in this way we cover also theories which are only spatially background-free, but use a preferred foliation.

The definition of a spatially background-free theory of gravity is based on a straightforward notion of equivalence of semiclassical configurations:

\begin{definition}[equivalence]
Assume the two configurations of one-particle semiclassical gravity $\gx{},\lx{}$ and $g'_{\mu\nu}\xt,\phi'\xt$ can be obtained from each other using a spatial coordinate transformations $x'=x'(x) = (\mathbf{x'}(\mathbf{x},t),t)$ via the usual hole construction. That means, the application of the coordinate transformation $x'(x)$ to the fields $g_{\mu\nu}\xt,\phi\xt$, where $g_{\mu\nu}\xt$ is transformed like a metric, and $\phi\xt$ as the square root of a density, gives $g'_{\mu\nu}(\mathbf{x'},t)$, $\phi'(\mathbf{x'},t)$, with $\mathbf{x'}$ instead of $\x$. Then the two configurations are said to be equivalent.
\end{definition}

Based on this notion of equivalence we can define now the meaning of background independence:
\begin{definition}[background independence]\label{def:nobackground}
A quantum theory of gravity is spatially background-free if equivalent semiclassical one-particle configurations $g_{\mu\nu}(\x),\phi(\x)$ and $g'_{\mu\nu}(\x),\phi'(\x)$ define (modulo the minimal uncertainties required by the uncertainty principle) the same quantum state \gk{}\lk{}.
\end{definition}
In particular, a background-free theory in the usual sense will be also spatially background-free.

Now we can formulate our main impossibility result:
\begin{theorem}\label{t:impossibility}
A viable quantum theory of gravity which fulfills postulate \ref{p:main} cannot be spatially background-free.
\end{theorem}

\begin{proof}
Let's consider two configurations $\gx{1},\lx{1}$ and $\gx{2},\lx{2}$. If there are no other than gravitational interactions and in the limit where gravitational interaction can be ignored we have $\abs{\pro}=1$. This limit can always be reached by moving the test particle away from the positions $x_i$ of the heavy particle. Thus, we can assume $\abs{\pro}\approx 1$.

Then let's assume that the wave functions $\phi_{i}$ have finite support $U$ in space for each moment of time $t$. This can be reached at least approximately, with a probability of the particles being outside $U$ being sufficiently small, by some independent external potential. The following argumentation works also in this approximate case. 

We consider now a spatial coordinate transformation $x'\xt = (\x'\xt,t)$ with the following properties:

1. It is trivial ($\x'\xt=\x$) for $t\le t_0$ and for $\abs{\x}>C$ for some large enough $C$.

2. At times $t>t_1>t_0$ for all $\xt\in U$ we have $(\x'\xt,t)\not\in U$. That means, the image of $U$ is contained in a set $U'$ which has for $t>t_1$ no intersection with $U$.

For a spatially bounded $U$ and large enough $C$ it is always possible to construct such a transformation.

We apply now the hole construction as described in section \ref{s:hole} to $x'\xt$ to $\gx{1},\lx{1}$. In the first step, we have to compute the transformed functions ${g^1_{\mu\nu}}'(\x',t),\phi'_1(\x',t)$ in the new coordinates. In particular, the wave function transforms like a square root of a density, so that
\begin{equation}
 \phi'_1(\x'(\x,t),t) = n(\x,t) \phi_1(\x,t)
\end{equation} 
with a weight factor $n(\x,t)$ of an irrelevant form.\footnote{To obtain that $\overline{\phi}(\x)\phi(\x) d^3\x$ at a given time $t$ transforms like a measure one has to use $n(\x,t) = \sqrt{\det \mathcal{J}}^{-1}$ where $\mathcal{J}$ is the three-dimensional Jacobi matrix $\mathcal{J}^{i'}_i = \partial x^{i'}/\partial x^i$.}
Once $\phi_1\xt\neq 0$ only if $\xt\in U$, it follows that for $t>t_1$ $\phi'_1(\x',t)\neq 0$ only if $(\x',t)\not\in U$.
Then, following the hole argument, we replace the coordinate $(\x',t)$ by $\xt$ to obtain the modified field configuration ${g'}^1_{\mu\nu}(\x,t),\phi'_1(\x,t)$. Because of property (1) of the coordinate transformation this modified configuration ${g'}^1_{\mu\nu}(\x,t),\phi'_1(\x,t)$ fulfills the same initial and boundary conditions as the original configuration ${g}^1_{\mu\nu}(\x,t),\phi_1(\x,t)$, thus, the two configurations are equivalent. Once by assumption the theory is spatially background independent, the two configurations define the same state \gk{1}\lk{1}.

Now, because of the points \ref{i:sesqui}, \ref{i:exp} of postulate \ref{p:main} the observable $\oo$ may be computed at the final time $t_{out}>t_1$ as
\begin{equation}\nonumber 
\o = \int \overline{\lw{1}}\xt \lx{2} d^3x,
\end{equation}
using the wave function $\lx{1}$. But once the same state \gk{1}\lk{1} can be obtained also by the modified configuration ${g^1_{\mu\nu}}'(\x,t),\phi'_1(\x,t)$, $\oo$ may be computed also as
\begin{equation}\nonumber 
\o = \int \overline{\phi'_1}\xt \lx{2} d^3x.
\end{equation}
By assumption, the first formula gives approximately $1$. Instead, the second formula gives (approximately) $0$.  Indeed, \lx{1}\/ has its support in $U$, and because of $\abs{\pro}\approx 1$ this holds approximately for \lx{2}\/ too. But for $\xt\in U$ we have $\phi'_1\xt=0$. Thus, the product $\overline{\phi'_1}\xt \lx{2}$ is (approximately) zero everywhere. In more detail,
\begin{equation}
\begin{split}
 \ao &= \iabs{\int \overline{\phi'_1}\xt \lx{2} d^3x} \\
&\le \iabs{\int_U \overline{\phi'_1}\xt \lx{2} d^3x} + \iabs{\int_{\overline{U}} \overline{\phi'_1}\xt \lx{2} d^3x}\\
&\le \sqrt{\int_U \overline{\phi'_1}\xt \phi'_1\xt d^3x}\sqrt{\int_U \overline{\phi_2}\xt \phi_2\xt d^3x}\\
&\quad + \sqrt{\int_{\overline{U}} \overline{\phi'_1}\xt \phi'_1\xt d^3x}
 \sqrt{\int_{\overline{U}} \overline{\phi_2}\xt \phi_2\xt d^3x}\\
&\le \sqrt{\int_U \overline{\phi'_1}\xt \phi'_1\xt d^3x} + \sqrt{\int_{\overline{U}} \overline{\phi_2}\xt \phi_2\xt d^3x}
\end{split}
\end{equation} and, if the support of $\phi_2$ is in $U$ and the support of $\phi'_1$ in $\overline{U}$ above integrals are zero. But even if $U$ is only approximately the support of the $\phi_i$, the probability of being outside $U$ should be small, thus,
\begin{equation}
\int_{\overline{U}} \overline{\phi_2}\xt \phi_2\xt d^3x\le\epsilon\quad \text{ resp. }\quad \int_U \overline{\phi'_1}\xt \phi'_1\xt d^3x\le \epsilon,
\end{equation} 
thus, $\ao\le 2\sqrt{\epsilon}$ is approximately zero.

Thus, we have obtained two incompatible values for the same observable $\ao$. Even if we allow for some inaccuracies because of the approximations used, the approximate results remain incompatible: $1$ and $0$ are not simply two different values, but the maximal and minimal possible value for $\ao$, so that certainly $1 \not\approx 0$ for any meaningful notion of approximation.

But a viable theory should make unique predictions for all observables. Therefore a theory with the assumed properties is not viable. 
\end{proof}

It is worth to note that we have applied the hole construction only to one of the two metrics in the superpositional state. One could ask if it wouldn't be more appropriate to apply the same diffeomorphism to above metrics. This would correspond to the notion of c-covariance considered by Anandan \cite{Anandan}, to be distinguished from the stronger notion of q-covariance, where different diffeomorphisms can be applied to different metrics. But, in agreement with Anandan, it is the stronger notion of q-covariance which corresponds to background freedom in the sense of GR. In fact, the notion of ``the same diffeomorphism'' is simply not defined for different solutions of GR. To define it in a meaningful way, one needs -- a background. 

\section{Recovering the common background}

An impossibility theorem seems to be something negative: It is impossible for quantum gravity to extend a very nice and beautiful property of general relativity -- background independence -- into the quantum domain. Quantum gravity would have to be less symmetric and, therefore, less beautiful than expected by those who argue for a background-free theory of quantum gravity.

But we have to disagree here. In our opinion, the result is a positive one, if we look at it from another side -- from the point of view of the possibilities of observation. Theoreticians have focused their interest mainly on the loss of possibilities caused by quantum theory -- the uncertainty relations. But there is also another side: Quantum theory gives also new possibilities for measurements. Almost all of the impressive progress of the accuracy of our measurement devices is based on the use of various quantum effects. But the improvement of accuracy using interference effects is not all what quantum measurements have to offer. It gives also qualitatively new observables -- additional information which cannot be gained even in principle using only classical measurements.

A similar insight has been gained already by the Bohm-Aharonov effect: In a region (outside a toroidal solenoid) where the EM field strength $F_{\mu\nu}$ was identically zero it was nonetheless possible to measure a nontrivial EM effect -- a phase shift in an interference picture. This phase shift may be computed by an integral over a closed path of the gauge potential $A_{\mu}$. This is a strong hint that the true degrees of freedom of a gauge field are the gauge potentials $A_{\mu}$ and not the field strength  $F_{\mu\nu}$. Unfortunately, this argument is not decisive -- one cannot recover the gauge potential $A_{\mu}$ even if all the integrals $\oint A_{\mu}(x) dx^\mu$ over closed paths are known.

This leads to the question about the relation between the new quantum observables \oo\/ and the background. We have found that we need the background to compute the observable \oo. But is it possible to recover the background given all possible observables of type \oo? Or is there a similar difference as between the classical gauge potential $A_{\mu}$ and the observable integrals over closed paths $\oint A_{\mu}(x) dx^\mu$? This is answered by the following

\begin{theorem}\label{t:background}
Assume in a quantum theory of gravity the items \ref{i:first} -- \ref{i:exp} of postulate \ref{p:main} hold and there is no superselection rule which prevents superpositions of different semiclassical states. Then we can define a common background shared by all gravitational fields in this theory.
\end{theorem}

\begin{proof}
For an arbitrary given field \gx{} we consider its superposition with some fixed vacuum state $\eta_{\mu\nu}$. Because of postulate \ref{p:main} item \ref{i:exp}, observable \oo, defined for all possible experiments of this type, and, therefore, for all possible outcomes \lk{1}, \lk{2}, is sufficient to define the sesquilinear form on $H_\qk{g}\times H_\qk{\eta}$ completely. Because it is non-degenerated, it can be used to define an isomorphism $H_\qk{g}\cong H_\qk{\eta}$: Every element of $H_\qk{g}$ defines a linear functional on $H_\qk{\eta}$, thus, an element on this space. Given this isomorphism, the position measurement on the vacuum state, defined by some projector-valued measure on $H_\qk{\eta}$, defines uniquely a projector-valued measure on $H_\qk{g}$. This measurement on $H_\qk{g}$ defines the common background on \qk{g}.
\end{proof}

Here, we have not used the explicit form of the scalar product as postulated in item \ref{i:explicit} of postulate \ref{p:main}. The abstract form defined by item \ref{i:sesqui} was sufficient. But the explicit product form guarantees better properties:

\begin{theorem}\label{t:backgroundcommuting}
Assume in a quantum theory of gravity postulate \ref{p:main} holds. Then the measurement of the background position commutes with the measurement of position on the metric \gx{}.
\end{theorem}

\begin{proof}Indeed, the coordinate $\x$ used in item \ref{i:explicit} of postulate \ref{p:main} defines position on above metrics involved in the superpositional state. Thus, it defines a position measurement on the metric \gx{} as well as on the background $\eta_{\mu\nu}$.
\end{proof}

The construction of the background in the proof of theorem \ref{t:background} contains a certain freedom of choice -- the choice of the reference state $\eta_{\mu\nu}$ we have named ``vacuum''. But a different choice of the vacuum does not change much -- it simply corresponds to the freedom of choice of the coordinates of the common background. The identification of ``the same point'' in different gravitational fields will remain unchanged.

Of course, a single experiment gives always only one value \oo, while for the recovery of the background in our construction we need all the values \pro\/ for arbitrary wave functions \psn{1}, \psn{2}. But this situation is typical in physics: To specify whatever field configuration completely by measurement, we would need an infinity of measurements, moreover of infinite accuracy, which is impossible. The positivistic idea that all we use in physics has to be based on results of measurements is simply nonsensical. The purpose of the theorem is a different one: It shows that all the \emph{information} which is, in each peace of it, available in principle by measurement, is sufficient to recover the background. 

This distinguishes our quantum gravity observables from the observables of the Bohm-Aharonov effect. These do not allow to reconstruct the gauge potentials themself, even if we would know all results of all in principle possible experiments. The experiment gives  only the gauge-invariant integrals of the gauge potential over a closed path.

Note that to prove that the background is unobservable classically, we need the covariance of classical general relativity. In this sense, the insight of classical GR is far away from being lost -- it is what allows us to prove that \emph{quantum measurements are really more powerful} than classical measurements.

\section{Discussion}\label{s:discussion}

The background-freedom of GR is a really beautiful property, and the idea to extend this property into the quantum domain has its beauty and should not be given up too early. But those trying to find such a theory already know that this job is almost impossible, and that in particular the problem of defining the observables is very critical.\footnote{In particular Smolin \cite{Smolin} notes that ``\ldots one cannot define the physical observables of the theory without solving the dynamics'', and Thiemann \cite{Thiemann} writes ``\ldots one must find a complete set of Dirac observables (operators that leave the space of solutions invariant) which is an impossible task to achieve even in classical general relativity.''} Despite this, they seem quite optimistic. This suggests that they will hardly give up reading about our impossibility theorem.

But are there any objections against our impossibility theorem which may justify some hope? There is of course the general hope that impossibility theorems in physics often do not hold their promises. There are lot's of examples of impossibility theorems which have later appeared to be based on unreasonable assumptions.\footnote{The classical example are the various impossibility theorem for hidden variable theories, starting with von Neumann's, where one ``impossible'' theory is explicitly known today as de Broglie-Bohm pilot wave theory.}

This is not much, only a quite abstract hope. But it is sufficient to motivate us to consider possible objections in a very careful way.

\subsection{Why we consider approximations}\label{a:whyApproximations}

One may think that it would be better to consider properties of full quantum gravity directly. Using approximations is only the cause of inaccuracies and uncertainties. The results become not only inaccurate -- we also usually do not have information about the size of the error made in the approximation. Thus, one should use approximations only if necessary, and use exact properties of the full theory whenever possible.

This is the normal thinking of every working scientist. It is the correct way to think in almost all the problems scientists have to do -- from developing new theories to deriving predictions from existing theories. And it is so obvious that it becomes part of the scientific intuition.

But there is an exception, and what we do here -- to prove an impossibility theorem for theories -- is this exception.
The logic of impossibility theorems is reverted. The weaker the assumptions we have to make about a theory, the stronger the resulting impossibility theorem. Indeed, an impossibility theorem becomes stronger if more theories are covered, shown to be impossible. And if we make weaker assumptions, more theories fulfill these assumptions and are, therefore, covered by the impossibility theorem. A logical triviality.

But if we make assumptions about an approximation of a theory, instead of making the same assumptions about the theory itself, we make a weaker assumption. It covers not only the theories which exactly fulfill this assumption, but also all those which fulfill this assumption only approximately, all those where exists some domain of applicability of the particular approximation. This is a much larger class of theories. And, therefore, the impossibility theorem for this much larger class of theories is a much stronger result.

We conclude that, in an impossibility theorem, it is preferable to fix not properties of the theory itself, but, instead, properties of approximations. This preference for approximations seems to be (judging from some reactions to earlier versions of this paper) highly counterintuitive. And this is quite natural -- the usual preference for exact theories is correct in almost all situations and therefore easily becomes part of the intuition.

There is also another aspect of this point. We usually use approximations because they are simpler mathematically, which allows to solve some mathematical problems which cannot be solved in the exact theory. This is an effect we use in this paper too -- our preference for approximations simplifies the mathematics we have to use. But it is, again, highly counterintuitive. The one who is able to manage more complex mathematics is intuitively the better scientist. But here we prefer to use the simpler mathematics of approximations and claim that the result becomes even stronger.

Nonetheless, this counterintuitive claim is correct. It is a straightforward consquence of the logic of impossibility proofs. To help the intuition to feel that this is correct, it is useful to think about another problem: What is necessary to circumvent a given impossibility theorem.

Assume that the impossibility theorem requires an exact property of the theory. In this case, to circumvent the impossibility theorem, one has to develop a theory where this property does not hold exactly. If it holds only approximately, the impossibility theorem is already no longer applicable to the new theory.

The situation is much more difficult if we assume, in the impossibility theorem, that the property holds only in some approximation. In this case, the simple way to circumvent the problem no longer helps. We have to invent a more strange theory, one where the property does \emph{not even hold approximately}. 

\subsection{Why we prefer weaker approximations}\label{a:whyWeakerApproximations}

Similar counterintuitive conclusions follow if we consider the question which approximation we have to prefer. Standard scientific intuition gives a clear answer: The more accurate, better approximation is preferable. It gives higher accuracy, and less uncertainty. But, again, the reverted logic of impossibility proofs give another answer: An impossibility theorem becomes stronger, if it fixes properties of weaker, less accurate approximations.

The logic behind this is, of course, the same. Theories which share a weak approximation may already differ in better approximations. Thus, the class of theories which share a given better approximation is smaller than the class of theories which shares a weaker approximation. An impossibility theorem based on the properties of the better approximation will cover only this smaller class of theories. But an impossibility theorem becomes stronger if it covers more theories.

To support our intuition let's consider again the problems faced by somebody who wants to circumvent the impossibility theorem. The mathematics of the weaker approximation is simpler, there are, for example, less parameters in a lower order approximation than in a higher order approximation. As a consequence, there is also less freedom to change the approximation to circumvent the impossibility theorem.

But there is also another aspect of this. It is well-known and obvious that it is easier to test experimentally lower order effects.  Thus, the parameters of lower order, weaker approximations are much more restricted by experiments. This additionally restricts the possibilities to circumvent an impossibility theorem based on the lower approximation.

For our quantum gravity considerations this does not apply directly -- the effects are far away from our experimental possibilities. But it applies in some indirect way. There is also another domain -- the domain where we cannot do experiments, but where we nonetheless have good intuitions, expectations about what will happen. And this domain of reasonable expectations has similar properties: Lower order, weaker approximations are more restricted by our reasonable expectations than higher order approximations.

And this is already an important point. While we have no nontrivial experiments with decoherence caused by gravity, we have reasonable expectations about it. They are extrapolations from decoherence caused by other interactions, supported by the consideration of Newtonian gravity and, last but not least, by the trivial observational support that, as long as gravity is too weak, it does not lead to decoherence. About other quantum gravity effects (say, superpositions of black holes, topological foam, effects near singularities) we have much less reasonable expectations. But these reasonable expectations restrict only the weakest possible approximations of quantum gravity. They don't tell us anything about full quantum gravity or higher order approximations.

Thus, it is not only preferable for impossibility theorems to fix only properties of approximations instead of properties of the exact theory.  It is also preferable to use weaker approximation. Impossibility theorems based on weaker approximations are stronger, cover more theories, and it is much more difficult to circumvent them.

Nonetheless, given the counterintuitive character of this logic of impossibility theorems, it seems useful to evaluate the particular instances where we apply such a preference for weaker approximations. 

\subsection{What about strong gravitational fields?}\label{a:strong}

One may ask what our considerations can tell us about strong gravitational fields.

The answer is simple: Nothing. Strong gravitational fields are irrelevant for our impossibility argument. A viable theory of gravity should be able to describe all gravitational fields, weak ones as well as strong ones. We show that a background-free theory is already unable to handle weak gravitational fields appropriately. This is sufficient to prove that the theory is not viable as a general theory.

A referee found this consideration ``suspect'', and explained that one has to ``expect problems to come from strong gravitational fields or possibly closed universes rather than just from weak fields''.  But this is only another case of the inverted logic of impossibility theorems. For those who try to find a quantum theory of gravity, the strong fields are much more problematic than the weak fields, and a result which applies even to strong fields would have much higher value. But we prove an impossibility result. And if we can prove that a certain idea fails already for weak fields, this  gives a much stronger result.

Of course, the problems related with strong fields or possibly closed universes can be expected to be greater than those for weak fields. And for the quantum gravity researcher it will be quite natural to care more about the more serious problems. If one finds a way to solve the problems of strong fields, one can hope that the problems of weak fields will be solved by the way too. But what would be the gain of a solution of a problem for weak fields if the problems for strong fields are much greater? A reasonable logic. 

But it may be misleading. In fact, the straightforward solution of the problem presented here -- the introduction of a background -- solves a lot of problems with strong fields. In particular we no longer have to consider fields with nontrivial topologies, and there will be no ``topological foam''.

And, more important, what counts here are not the strategies of quantum gravity researchers -- it remains their freedom to decide which problems are worth to be tackled first. The question is only if the problems with strong fields lead to some argument against the theorem presented here. This is certainly not the case. If the problems with weak fields are proven to be unsolvable, even greater problems with strong fields will certainly not save the day. 

\subsection{What about different foliations of spacetime?}

We consider everything only in a given foliation. But a background-free theory clearly should not have a preferred foliation. So, one may ask, what about different foliations? In particular, what about the problems with unitarity related with different foliations of spacetime already in a Minkowski context, as described in \cite{Torre}? 

Again, our answer is simple: We don't have to care about this. If the final quantum theory of gravity is able to handle different foliations of spacetime, or if it solves the ``problem of time'' in a quite trivial way, using some preferred foliation, does not matter at all. In any case, the theory should be able to handle appropriately at least one foliation of spacetime. Our theorem shows that even spatially background-free theories are not viable, thus, a background-free theory would not be viable even if restricted to a single foliation.

This is yet another case of the same reverted logic of impossibility theorems: The much greater problems related with different foliations of spacetimes will certainly not save the day for theories which already fail to manage a single preferred foliation correctly. 

\subsection{Maybe the formula for the scalar product is false?}

The formula for the scalar product \eqref{d:product} is of central importance for our considerations. But it is a formula supposed to be valid in a domain -- superpositions of relativistic gravitational fields -- where we have no established theory. So one may argue that this formula may be invalid in this domain.

But there is good evidence for this formula. First, it holds in the non-relativistic limit. This is simply many-particle \Sch theory, a well-known and well-established theory, which is in parts even supported by experiment \cite{neutrons}. In this theory, the source particle and the lightweight particle are two particles which interact via the Newtonian potential. Once the source particle is assumed to be much heavier, and to be in states \hk{2} and \hk{1} strongly localized near the slits $x_i$, one can use even the one-particle theory with the Newtonian potential created by the source particle in $\x_i$ as an external field to compute $\psi_{1/2}\xt$ for the initial value $\psi_{1/2}(\x,t_0) = \psi_0(\x)$:
\begin{equation}
	i\partial_t \psi_{1/2}(\x,t) = (-\frac{1}{2m} \Delta - \frac{mM}{|\x-\x_{1/2}|})\psi_{1/2}(\x,t)
	\label{e:Schroedinger}
\end{equation}
The resulting functions $\psi_{1/2}\xt$ define two approximate solutions of the two-particle problem
\begin{equation}
 \Psi_{1/2}(\x_1,\x_2,t) = \delta(\x_1-x_{1/2})\psi_{1/2}(\x_2,t).
\end{equation}
The derivation given in the proof of theorem \ref{t:measurement} is then valid in two-particle theory, with the scalar product for the lightweight particle is defined by \eqref{d:product}. So, in this limit, \eqref{d:product} holds.

But there is also another limit where \eqref{d:product} holds: Semiclassical theory. This theory is applicable if the gravitational fields \gx{1}\/ and \gx{2}\/ are equal. In this case, the gravitational interaction is irrelevant. But the formula nonetheless holds. If the source particle and the lightweight particle interact in other ways, for example electromagnetically, the formula gives nontrivial predictions for electromagnetic interference effects in external gravitational fields. Experiments in this domain are standard, and, given that gravitational redshift is a relativistic effect, relativistic gravity is required.

Thus, the formula holds in two different limits: The non-relativistic as well as the semiclassical one. 

Third, we have the proof of theorem \ref{t:measurement}. What is used there are only quite general quantum principles. Thus, even in quite different situations (quite different environments) the result has to be of the same general type, namely a scalar product.

Moreover, we do not need the formula nor for strong gravitational field, nor as an exact formula even for weak fields. All we need is an \emph{approximate} formula for weak relativistic fields. An approximate formula is sufficient because for the different states constructed in our hole argument the value of \abs{\oo} varies from the maximal possible value $1$ to the minimal $0$, which is clearly too much for an approximation error.

But even in the worst case, even if we do not have the explicit formula \eqref{d:product}, theorem \ref{t:background} gives us an observable background. This background may not have the nice properties following from the theorem \ref{t:backgroundcommuting}, but a background is a background is a background, and our goal was to prove that no background-free theory is viable.

\subsection{Maybe Newton-Cartan is more appropriate than Newton?}

In the previous point, we have used Newtonian theory as the non-relativistic limit to justify the scalar product formula. But is this justified? It is known that Newton-Cartan theory is the correct $c\to\infty$ limit of GR, and this theory is already quantized \cite{Christian}.

But the Newtonian theory as a limit is not wrong -- we simply need additional restrictions to obtain it. We have preferred it because of its simplicity, in particular in application to our two-particle problem. Moreover, as we have already mentioned, we do not have to use the most advanced approximation if we want to prove an impossibility theorem: The weaker the approximation, the larger the class of theories covered.

\subsection{Maybe in full quantum gravity it is meaningless to talk about interacting particles?}

In our thought experiment, we use particles as if they were simply quantum billiard balls. But already in semiclassical field theory we have to handle particle creation and destruction by the gravitational field, and it seems much more appropriate to consider particles only as phenomenological, derived objects: The fields seem to be more fundamental. In full quantum gravity the situation may be even worse. It may become completely meaningless to talk about particles -- we don't know yet. So, one may argue that our thought experiment becomes meaningless in full quantum gravity.

But even in full quantum gravity there will be a domain where particles are a good approximation. This is certainly so for the physical situations which we observe today. And this will remain to be true in the domain of non-relativistic, Newtonian quantum gravity. In this domain, our thought experiment clearly makes sense.

Moreover, the theorem \ref{t:measurement} is not restricted at all to the consideration of a single particle, not even with some particles. Only the formula \eqref{d:product} depends on this. All we need for the other points of theorem \ref{t:measurement} is that it interacts with some other system, which has a separate Hilbert space with a scalar product in it.

\subsection{Maybe the experiment makes sense only for such weak fields that the Newtonian limit is sufficient to handle it?}

If one accepts that the experiment makes sense for some weak fields, where Newtonian gravity is able to handle it approximately, one could try to argue the other way around:  Maybe the domain where the experiment makes sense is sufficiently small, essentially containing only the Newtonian limit?  In this case, one could argue that in this limit we can use the fixed background of Newtonian theory to predict the result of the experiment.

There is much to answer. First, it would be doubtful that a theory which uses the background in the Newtonian limit is background-free. It is known, in particular, that the true $c\to\infty$ limit of GR is not Newtonian theory, but Newton-Cartan theory, which leads to Newtonian theory only if one imposes an additional condition \cite{Christian}. What will be the true $c\to\infty$ limit of a background-free full quantum gravity? It seems reasonable to guess that it will be a background-free quantization of Newton-Cartan theory, where some principle of background-independence similar to definition \ref{def:nobackground} holds. This theory would face the same impossibility problem as a background-free GR quantization.

Of course, in Newton-Cartan theory it is natural to use the background structure which exists in this theory for quantization. But one should not expect that this quantum theory appears as the limit of a background-free theory. It is much more reasonable to expect this theory as the limit of a GR quantization which uses some background too, for example, of a quantization of GR in harmonic gauge.

Second, the background exists only in the limit.  It does not exist for weakly relativistic fields.  However weak the relativistic effects may be in a certain situation, there is always some level of accuracy such that the Newtonian limit is unable to make a sufficiently accurate prediction. For this level of accuracy, the Newtonian limit itself already does not provide any fixed background. Of course, one would need only an approximate background, but this approximate background should be already a structure of relativistic gravity, different from the background in the limit.

Third, it seems unreasonable to expect that the experiment makes sense only in an extremely small environment of the Newtonian limit. Especially if we look at theorem \ref{t:measurement} and its proof, we find that the thing our source particle interacts with does not have to be specified at all. All we need is the quite general principle of quantum theory that the combination of a system and its environment has to be described by a tensor product of the two Hilbert spaces, with a scalar product defined in each of them. This general quantum principle works quite fine for highly relativistic particles used in current particle accelerator experiments. Thus, relativistic velocities do not seem to be problematic in any way for this principle. Thus, it seems reasonable to expect that we will obtain some observable complex value \oo\/ even if the lightweight particle appears highly relativistic, if there will be a lot of relativistic particle creation and destruction, and so on.

Last but not least, even if we doubt that such fundamental principles of quantum theory as those used in the proof of theorem  \ref{t:measurement} survive in full quantum gravity, it is hard to expect that they start to fail already for weak to moderate relativistic fields. But let's consider this in some more detail.

\subsection{Maybe in quantum gravity there are no tensor product structures?}

In the proof of theorem \ref{t:measurement} we have used a tensor product decomposition $\H_S\otimes\H_E$ of the full Hilbert space into that of the system (the heavy particle) and its environment (the lightweight test particle). Maybe this is the weak place and there is no such decomposition in quantum gravity?  In this case, the proof of theorem \ref{t:measurement} fails and there may be something wrong with the thought experiment.

First we object that even if this would happen, it would not mean that theorem \ref{t:measurement} becomes automatically wrong. Nor the assumptions, nor the claims of the theorem refer to the tensor product structure. We do not have a proposal for such a more complicate structure and therefore cannot evaluate here the properties of such a structure, so this remains speculative. Nonetheless, let's note here that we do not need much: If the two basic states \hk{i}\/ remain unchanged by the interaction, it seems quite reasonable that every theory worth to be named ``quantum theory'' would lead to a very close restrictions for superpositions of these states too. And all we need to prove the theorem are the following properties of superpositions of the \hk{i}:
\begin{itemize}
 \item The result will be a density matrix $\sum \rho_{ij} \hk{i}\hb{j}$ on the subspace generated by the \hk{i},
 \item and the values of the diagonal elements $\rho_{ii}$ remain unchanged. 
\end{itemize}
These are assumptions which in no way refer to the tensor product structure, and which make a lot of sense independently of any assumptions about a tensor product structure. So they may remain valid even if the tensor product structure will be replaced by some more complicate structure. Or, if we look at it from the other side: A theory which violates these principles seems even more suspect than a theory which replaces the tensor product structure by something more complicate.

But let's nonetheless consider the worst case scenario that this doesn't help, and even those conditions are violated by some theory of quantum gravity. Then the natural question is about the domain where the theorem remains valid as an approximation. 

This domain is not empty, it contains everything accessible to observation today. As a consequence it already includes interactions which are many orders stronger than gravitational interaction, already includes strong special-relativistic effects, and, given that in experiments about gravitational redshifts quantum effects are used, already includes even some gravitational effects.

Then, if the theory in question has Newtonian gravity as its non-relativistic limit, this domain also contains non-relativistic gravity. But in Newtonian gravity the effect may be already large. That means, the domain of applicability also contains situations where our thought experiment gives non-trivial results.

But once nor gravity in itself, nor special relativity in itself, nor the much stronger fields we observe in real experiments in itself, are in conflict with the theorem, one would expect that there is also a rather large domain where all three complications are present but the theorem nonetheless remains applicable as an approximation.

A quite different speculation leads to a similar result. Different quantum effects will be important in different regions of the parameter space. Decoherence becomes important whenever the gravitational fields of the particles are strong enough to lead to a quite minimal modification of the states of a test particle. The minor modifications of the states of test particles which are sufficient to get the maximal effect of complete decoherence are probably undetectably small with any other method, in particular if one remembers that even if the interaction with one particle does not lead to observable decoherence, the interaction with many of them may nonetheless lead to complete decoherence. Roughly speaking, if gravity is too weak to cause decoherence, quantum gravity is not yet relevant at all. And therefore it seems reasonable to expect that the more non-trivial quantum effects, the ones related with modifications of fundamental quantum principles, are important only in a much smaller region of much stronger, much more special fields.

Thus, it is reasonable to expect that even in theories where some fundamental quantum principles no longer hold, there will be a sufficiently large domain where the decoherence is the most important quantum gravity effect, and where the theorem is at least a good approximation.

But this would be already sufficient for us. Because, if we have such a domain where the experiment leads to nontrivial results, then the theory has to predict these results at least approximately.  The point of the impossibility theorem was that the result is completely undefined.

This last point remains valid even if theorem \ref{t:measurement} fails completely -- the experiment will give some result. Whatever it is -- it has to be predicted by the theory. And we have yet no idea how to do this in a background-free theory.

\subsection{Maybe one has to consider the full experiment?}

There is some standard way to describe the double slit experiment, with the particle starting from one starting point and ending at some other point at the screen. One could reasonably hope that a consideration which includes this full experiment does not require the consideration of any superpositions of gravitational fields:  We do not have to consider the intermediate states, we only have to compute the amplitudes between initial and final states.

But this simple picture of the double slit experiment is not a necessity. Instead, it is an idealization. A more general quantum experiment starts with states which are already superpositional states, and never have been localized in one point. In particular, the superpositional state may be the ground state of some appropriate double slit potential. Now, ground states may be prepared starting with an arbitrary state and waiting long enough for the energy being radiated away. In the same way, one may measure if the particle has remained in the ground state after the interaction: We look if the state radiates the energy which has been obtained during the interaction.

Then, we do not need a description of a particular experiment, but a general theory which allows to describe them all. But at every finite moment of time there will be not only semiclassical states, but also their superpositions. Thus, one would need a theory which is quite different from quantum theory: Possibly without any information about finite times, like an S-matrix theory, possibly without superpositions as legitimate states.

\subsection{But maybe the true theory of quantum gravity will be something like an S-matrix theory?}

In quantum field theory, we obtain physical predictions by computing the S-matrix. How to compute predictions beyond the S-matrix, for finite distances, in a gauge-invariant way is not clear. But this is widely considered as irrelevant -- all what we can compare with observations is the S-matrix.

In this situation, one may expect that in quantum gravity all we can compute is some analogon of the S-matrix as well. Taken together with some positivistic philosophy, it could be argued that some S-matrix-like theory of quantum gravity would be sufficient. For such a theory, it is not clear if our thought experiment makes sense -- we have used in our argument bounded states at finite times. Thus, one may hope that restriction to some S-matrix-like theory allows to avoid this problem.

Such a line of argumentation should be clearly rejected: The S-matrix is sufficient only FAPP (for all practical purposes).  But FAPP we simply don't need quantum gravity.  We need quantum gravity for theoretical reasons:  The true, final theory of the universe should be able to handle effects of quantum gravity, even if we, as human beings, are unable to test these predictions. A theory which does not cover quantum gravity is, therefore, insufficient theoretically, even if it covers, like semiclassical QFT on a GR background, all we can observe.

But this same line of argumentation which makes full quantum gravity interesting for us also shows that an S-matrix-like theory, which would be unable to make predictions for finite distances, is insufficient too. Our human experience is bounded to finite distances, thus, the infinite limits gives by the S-matrix are not what we really observe, but only an approximation of our observations. If this approximation is sufficient FAPP or not is completely irrelevant for its \emph{theoretical} status as an approximation. The true theory of the universe should be able to predict our observations, at least in principle, with an accuracy higher than that given by the S-matrix.

\subsection{The relationalist argument}

One reason for the attractiveness of the background-independent approach is its philosophical background. While many scientists consider philosophy as off-topic or irrelevant in scientific articles, others (including the author) think otherwise, and articles in favour of background freedom, like \cite{Smolin}, use philosophical arguments at a prominent place. So we think these arguments are worth to be considered here too.

The philosophical argumentation in favour of a relational, background-free theory goes back to the position of Leibniz, who has proposed arguments for a relational view, against the absolute notion of space and time proposed by Newton. A nice introduction, from point of view of the modern background-independent approach, can be found in \cite{Smolin}:

\begin{quote}
``Leibniz's argument for relationalism was based on two principles, which have been the focus of many books and papers by philosophers to the present day. The \emph{principle of sufficient reason} states that it must be possible to give a rational justification for every choice made in the description of nature. \ldots A theory that begins with the choice of a background geometry, among many equally consistent choices, violates this principle.\ldots

One way to formulate the argument against background spacetime is through a second principle of Leibniz, \emph{the identity of the indiscernible}. This states that any two entities which share the same properties are to be identified. Leibniz argues that were this not the case, the first principle would be violated, as there would be a distinction between two entities in nature without a rational basis. If there is no experiment that could tell the difference between the state in which the universe is here, and the state in which it is translated 10 feet to the left, they cannot be distinguished. The principle says that they must then be identified. In modern terms, this is something like saying that a cosmological theory should not have global symmetries, for they generate motions and charges that could only be measured by an observer at infinity, who is hence not part of the universe.''
\end{quote}

From point of view of the Popperian scientific method \cite{PopperCR, PopperLSD} relationalism has to be rejected as a variant of positivism, based on the priority of observation: The observation that some entities are indiscernible requires their identification in the theory. Instead, Popperian science is based on the priority of theory: Theories are free guesses about Nature, and they are, in particular, free to postulate differences between indiscernibles if this gives some other advantages. But this rejection of relationalism in principle does not diminish the heuristic value of its principles. While we are free to postulate differences between indiscernibles, we have to take into account Ockham's razor: Without good reasons to introduce  differences between indiscernibles we should nonetheless identify them.

Therefore it is important to recognize that our thought experiment invalidates the relationalistic argumentation against the background. First, the new \emph{observable} defines a \emph{sufficient reason} for the introduction of a common background: The background solves the problem of computing a prediction for the new observable, and there is no obvious alternative way to solve this problem. It is also no longer possible to apply the principle of ``identity of the indiscernible'' against the background. Indeed, superpositional states with different values for \pro\/ are no longer indiscernible. Last but not least, the background becomes itself a \emph{relational} object: As constructed here, the background defines a \emph{relation} between physical states -- two gravitational fields which are part of one superpositional state.

\section{Consequences for classical gravity}

We conclude that a background-free theory of quantum gravity is not viable. This has also consequences for classical gravity, because one thing we know for sure: That classical gravity has to be the $\hbar\to 0$ limit of quantum gravity. Once quantum gravity has a background, then there will be such a background in classical gravity too.

Thus, even if GR seems adequate for classical gravity, we have to incorporate a background into classical gravity. In particular, we need an equation for the background. Fortunately, the choice is simple -- there is a nice candidate preferred by its beauty:  The harmonic coordinate condition
\begin{equation}\label{e:harmonic}
\pd_\mu(g^{\mu\nu}\sqrt{-g}) = \square X^\nu = 0
\end{equation} 
is not only very beautiful condition in itself, but also simplifies the Einstein equations essentially. Harmonic coordinates have been used to prove local existence and uniqueness results for general relativity \cite{Choquet-Bruhat,Choquet-Bruhat1}, so that there is no non-uniqueness and no hole problem in GR in harmonic coordinates.

The identification of the background with harmonic coordinates already leads to some interesting modifications of GR: The topology of the solutions has to be trivial, we have a new notion of completeness (a solution is complete if it is defined for all values $-\infty<X^\mu<\infty$ of the harmonic coordinates $X^\mu$, while the metric $g_{\mu\nu}(x)$ no longer has to be geodesically complete), and we have a symmetry preference for the flat FRW universe (it is the only homogeneous universe).

Further modifications appear if we don't want to add the harmonic condition as an additional external equation, but want to obtain it as an Euler-Lagrange equation. The natural way to do this is to add a covariance-breaking term
\begin{equation}
 L_{GR} \to L_{GR} + \gamma_{\alpha\beta}g^{\mu\nu}X^\alpha_{,\mu}X^\beta_{,\nu}\sqrt{-g}
\end{equation}
for some constants $\gamma_{\alpha\beta}$ so that the harmonic condition becomes an Euler-Lagrange equation
\begin{equation}
 \frac{\delta S}{\delta X^\alpha} = \gamma_{\alpha\beta}\square X^\beta.
\end{equation}
This leads to additional terms in the Einstein equations. While we obtain the Einstein equations in the natural limit $\gamma_{\alpha\beta}\to 0$, this leads to even more qualitative modifications of GR even for arbitrary small values $\gamma_{\alpha\beta}$: Gravitons obtain a mass, we obtain stable ``frozen stars'' instead of black holes, and a ``big bounce'' instead of a big bang, and without the GR big bang and black hole singularities \cite{Logunov1, Logunov, glet}.\footnote{
Once a classically unobservable space-time background is introduced anyway, this background can be used to solve the ``problem of time'' of canonical quantum theory too -- in the most trivial way of introducing a preferred time. The corresponding ADM decomposition allows a condensed matter interpretation of gravity in terms of density, velocity and stress tensor so that the harmonic condition transforms into continuity and Euler equations \cite{glet,clm}. Because we know how to quantize condensed matter theories, such an interpretation may be useful for quantization. A preferred time may be useful for other problems too, in particular it may be used in realistic hidden variable interpretations of quantum theory like de Broglie-Bohm pilot wave theory, or for the condensed matter interpretation of the SM fermions and gauge fields proposed in \cite{clm}.
}

\section{Conclusion}

We have proven that a background-free quantum theory of gravity is not viable, is unable to predict the result of simple quantum experiments.

This is not simply yet another item on the long list of problems of ``we don't know what to do'' type on the way to such a theory. We know what to do. The background has been found to be observable in quantum theory. All the information contained in the background is in principle accessible by experiments of the type proposed in this paper. So, the solution is straightforward -- one has to introduce a common background into the theory of gravity.

This result is not only relevant for quantum gravity -- we have to introduce the background into classical gravity too, which leads to important modifications.

The experiment considered in this paper measures a weak field effect, one of the weakest quantum gravity effects imaginable. But the solution of the problem -- the introduction of a background -- simplifies quantization also for strong fields because we do no longer have to speculate about different topologies, wormholes, or topological foam.

Let's conclude with a remarkable observation: As in this case, as in the case of the Bohm-Aharonov effect the objects which allow to compute the new quantum observables -- the background as well as the gauge potential -- have been known long before quantum theory. This is a lecture about the remarkable power of mathematical simplicity and beauty, and the counterproductiveness of the positivistic rejection of unobservables.

\end{document}